# Gestion du déploiement de composants sur réseau P2P

Stéphane Frénot [1]

*INRIA ARES, Laboratoire CITI*
*Bat. Léonard de Vinci*
*69621 Villeurbanne Cedex*

*stephane.frenot@insa-lyon.fr*

*ABSTRACT: The deployment of component-based applications relies on a centralized directory to store the components. This paper describes an approach to distribute software components to be deployed on a set of peers of a peer to peer network in order to exploit some associated characteristics (load balancing, fault-tolerance, self-organisation). The proposed architecture is situated in the context of OSGI application deployment management. The software components (bundles) are distributed among a set of nodes participating in the execution of services. When a node wants to install a component which is not deployed locally, the component is looked for and installed using a p2p network.*

*RÉSUMÉ: Le déploiement d'applications à composants repose sur une approche d'annuaire centralisé de stockage des composants. Cet article décrit une approche pour distribuer les composants logiciels à déployer sur un ensemble de nœuds d'un réseau pair-à-pair afin de pouvoir exploiter certaines caractéristiques associées (équilibrage de charge, tolérance de panne, auto-organisation). L'architecture proposée entre dans le cadre de la gestion du déploiement d'applications sur le modèle OSGi. Les composants logiciels (ou bundles) sont répartis à travers un ensemble de nœuds participant à l'exécution de services. Lorsqu'un nœud veut installer un composant et si celui-ci n'est pas encore déployé localement, il est recherché et installé en utilisant un réseau p2p.*

MOTS-CLÉS: *pair-à-pair, déploiement, OSGi*

---

**1. Introduction**

L'installation de systèmes à base de composants logiciels nécessite une infrastructure de distribution et d'indexation de ces composants. Dans le monde des distributions linux, l'installation des *packages* de fonctionnement consiste à télécharger des *packages*, des archives rpm ou des archives tgz à partir de sites principaux ou miroirs, puis à les installer localement sur la machine cliente. La mise à disposition des composants logiciels pour ces plates-formes repose sur un serveur http qui centralise l'indexation des composants. Quand les charges augmentent, il est possible de mettre en place des miroirs permettant de gérer l'équilibrage de charge en terme de requêtes utilisateurs.

Dans cet article nous décrivons une mise en œuvre d'une infrastructure pair-à-pair (p2p) pour la gestion de composants logiciels "installables". Cette architecture décentralisée et distribuée s'abstrait de la notion de serveurs de distribution en répartissant les composants à installer sur l'ensemble des nœuds du réseau. De plus les composants les moins utilisés sont moins répliqués, ce qui permet d'avoir une auto-adaptation du réseau de déploiement.

Nous proposons une approche pour la gestion du déploiement de composants dans le monde OSGi qui exploite un sous-ensemble de nœuds pour l'indexation et la mise à disposition des composants.

Une première partie décrit le principe des plates-formes à composants et présente une structuration des réseaux p2p. Puis, après avoir décrit les fonctionnalités désirées pour notre système de déploiement, la quatrième partie propose une approche pour la gestion du déploiement fondée sur une infrastructure de nœuds organisés en réseau p2p. Enfin, en conclusion, nous discuterons de certains points clés liés à notre approche.

**2. Déploiement de composants et réseau pair-à-pair**

**2.1 Plates-formes à composants et position du problème**

*2.1.1 Présentation générale*

Une plate-forme à composants telle que nous la définissons dans cet article, est une infrastructure permettant la gestion du cycle de vie d'applications logicielles. Les applications logicielles sont récupérées à partir d'un site de dépôt sur l'Internet, puis sont déployées localement sur la plate-forme. Dans ce contexte, la notion de composant logiciel est l'unité de packaging et de transport de l'application. Le contenu de ce composant est décrit par un fichier appelé descripteur. Cette notion de composant s'applique dans de nombreuses approches. On peut citer à titre d'exemple : les composants dans la spécification J2EE (ejb, war, ear), les *packages* d'installation dans le monde linux (rpm, deb, ebuild) ou encore sur les conteneurs d'applications (sar phoenix, ou *bundles* OSGi). Si chaque plate-forme définit sa propre structure de *packaging* et de déploiement, on y retrouve facilement les concepts suivants :

- **Archive de transport** : L'ensemble des fichiers du composant sont regroupés dans une archive de transport éventuellement compressée (tgz, zip…).

- **Descripteur de déploiement** : fichier de description de l'archive. Le format peut être un shell d'exécution comme dans les ebuild de gentoo [GEN], un format xml (ejb) ou encore un fichier *manifest* java (OSGi).

- **Dépendance de composants** : il est possible d'exprimer des interdépendances entre composants (ou autres concepts). Ces dépendances permettent de contrôler si le package est installable ou non.

- **Cycle de vie du composant** : la plate-forme à composants gère le cycle de vie du composant (installé/démarré/arrêté/supprimé…), et chaque plate-forme propose un cycle de vie plus ou moins complexe.

Dans notre architecture nous nous sommes focalisé sur la problématique de mise à disposition des composants dans le cadre d'OSGi. Dans la section suivante nous décrivons plus précisément le modèle à composants de cette plate-forme.

*2.1.2 OSGi*

L'**O**pen **S**ervice **G**ateway **i**nitiative [OSG 02] est une proposition pour définir de façon standard la manière de gérer à distance des services et des périphériques matériels utilisés dans un environnement local. La spécification OSGi définit les API nécessaires pour pouvoir exécuter et gérer des services sur une passerelle. L'API OSGi repose sur la machine virtuelle et le langage Java. Elle spécifie les quatre concepts suivants :

**Le composant OSGi :** Le composant OSGi (l'unité de déploiement) est appelé *bundle*. Un *bundle* est une archive java au format jar, décrite par un fichier manifest spécifique à OSGi. De manière succincte, le composant contient et décrit les éléments suivants : la classe à exécuter au lancement du composant (start), les packages externes java nécessaires à son exécution (import), les packages java que le composant peut fournir (export).

**Le conteneur de composants**: c'est un démon java qui garantit l'exécution des différents composants hébergés. Il autorise et permet l'association entre des clients demandant l'accès à des services et des implantations de ces services. En résumé son rôle est d'enregistrer et de gérer localement l'activité de la plate-forme en termes de composants, de services et de packages java.

**Les services standards** : la spécification définit des services standards (http, logging, user admin, startlevel…) de la plate-forme. Ces services sont directement exploitables par d'autres services. Ainsi un nouveau service déployé peut toujours faire appel à un service web, pour fournir une interface web d'interaction.

**Un modèle de déploiement simple** : Le modèle de déploiement d'OSGi est volontairement simple et décentralisé (l'exécution des services se fait de manière locale à la passerelle mais l'administration est distante). Un composant peut être rapatrié à partir d'une URL distante puis installé et exécuté localement.

La figure 1 suivante replace le conteneur de service dans le contexte d'exploitation d'OSGi.

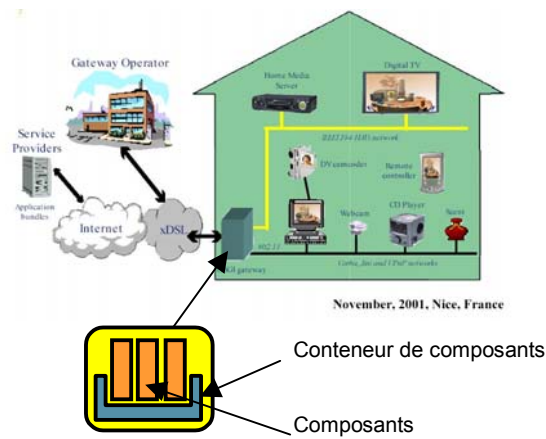

*Fig. 1 : Exemple d'exploitation d'OSGi*

Le conteneur de composants OSGi permet l'exploitation de *bundles*. Les *bundles* pilotent des périphériques dans le cadre du domicile d'un utilisateur. Ces *bundles* sont administrés à distance par un opérateur de passerelle et sont mis à disposition par des fournisseurs de services. Afin de fournir une gestion du déploiement différente à l'opérateur de la passerelle et aux fournisseurs de services, nous nous sommes intéressés aux approches liées aux réseaux pair-à-pair.

### 2.2 Réseaux Pair-à-Pair (P2P)

Les réseaux p2p sont très largement traités dans la littérature. Cette section ne se veut pas exhaustive mais extrait des différentes approches une structuration des réseaux afin de pouvoir choisir l'infrastructure la plus adaptée à la dissémination de composants.

Nous proposons de voir les réseaux p2p selon quatre axes : les réseaux à index, les réseaux à inondation, les réseaux à traînée et les tables de hash distribuées.

*2.2.1 Les réseaux à index*

Ces premiers réseaux correspondent à la vision « classique » du pair-à-pair. Le principe est d'avoir un réseau de pairs tous identiques reposant sur des index. Ces index sont centralisés ou distribués, et permettent l'indexation des ressources proposées par les différents pairs. Une ressource est indexée afin de maintenir l'association entre une ressource et un ensemble de pairs. Lorsqu'un pair recherche une ressource, il demande à l'indexeur de le mettre en relation avec un pair (ou un ensemble de pairs) possédant la ressource. Cette approche est largement la plus utilisée (Napster, Fastrack, Kazaa, Bittorent[2]) car elle permet de mettre à disposition du plus grand nombre des ressources personnelles. La simplicité pour diffuser les ressources et les récupérer en font des réseaux fortement adaptés à la diffusion de ressources diverses. Ces réseaux sont fragiles en deux points : la mise à disposition initiale de la ressource ne se fait que sur une seule machine et la fragilité du principe d'indexation qui repose sur un ensemble de nœuds pour identifier l'ensemble des ressources du système.

*2.2.2 Les réseaux à inondation et petits-mondes*

Un réseau à inondation ne possède pas d'index. Les ressources sont publiée localement sur une machine ; le système de recherche se fait par inondation de la communauté afin d'y retrouver la ressource [IVK 01]. L'inondation n'est que locale si la ressource est à proximité et s'étend globalement si la ressource n'est pas trouvée. Ce type de réseau répond à certaines propriétés spécifiques liées à la notion de « petits mondes » [WAT 03]. Ce type de réseau se focalise sur la recherche de ressources et présente donc des faiblesses concernant l'émission et la gestion du cycle de vie des ressources.

*2.2.3 Les réseaux à « traînée »*

Les réseaux à « traînée » cherchent à déposer la ressource initiale en plusieurs endroits afin de la rendre plus stable sur le réseau. Freenet [CLA 01] est le premier réseau de ce type. Cette approche est également adoptée dans le réseau à clé de hash *Pastry*[3] [ROW 02] où la ressource est routée à travers différents nœuds du réseau et où chaque hôte ainsi traversé reçoit une copie de l'association clé/localisation de la ressource. Lors de la recherche, le cheminement se fait en se rapprochant petit à petit de la source initiale de la ressource. Si le chemin permettant d'atteindre la ressource « croise » le chemin qui a permis de la déposer, la ressource est renvoyée par le nœud se trouvant à la jonction de ces deux chemins (cf. Figure 1).

---

[2] Bittorent est un réseau de pairs pour le transport de données volumineuses. Il ne possède pas d'index car ce n'est pas un problème du protocole. Cependant un index est nécessaire si on recherche une ressource qu'on ne connaît pas a priori. Les index des fichiers torrents sont classiquement maintenus sur des serveurs web.

[3] Malgré les caractéristiques de réplication de ce réseau, celui-ci est plus classiquement décrit dans la catégorie suivante correspondant aux réseaux à table de *hashage* distribuées.

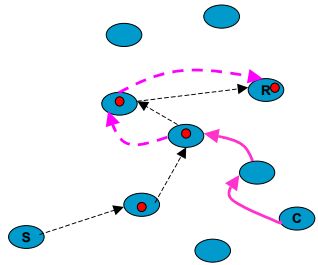

*Fig. 1 : routage dans le réseau pastry*

Le nœud source (S) dépose une ressource sur le réseau. Un chemin de routage (flèches droites) est calculé entre le nœud source et le nœud racine (R) ayant la clé d'identification la plus "proche" de la clé d'identification de la ressource. L'association clé/localisation est déposée sur tous les nœuds intermédiaires. Lorsqu'un client (C) recherche la ressource, il en calcule l'identifiant qui est routé de la même manière, par rapprochements successifs vers la racine de la ressource. Lorsque le chemin d'émission et le chemin de recherche se croisent la ressource est trouvée et renvoyée au client.

*2.2.4 Réseaux à tables de hash Distribuées (DHT)*

Un réseau pair-à-pair fondé sur les clés de *hash* distribuées réalise une correspondance 1-1 entre un identifiant et une valeur de *hash* [BRA 04]. Cette correspondance permet de placer directement une ressource dans le réseau de pairs.

La répartition des clés d'identifications est distribuée dans un anneau dans le cas de chord [STO 02], dans un hypercube dans le cas de *pastry* ou dans un espace euclidien virtuel dans le cas des réseaux CAN [RAT 01]. Dans chacun de ces réseaux les complexités algorithmiques liées à la recherche et à l'insertion des nœuds et des ressources varient et chaque réseau présente des avantages et/ou des inconvénients dans une de ces caractéristiques.

La section suivant essaye de décrire les besoins liés à la gestion du déploiement de composants logiciels.

**3 Fonctionnalités nécessaires à un réseau p2p de gestion du déploiement**

Notre architecture repose sur un ensemble de nœuds qui acceptent d'héberger des composants logiciels. Ces composants logiciels sont éventuellement récupérés par d'autres nœuds qui désirent les installer. Cette section a pour objectif de définir plus précisément les contraintes à résoudre si l'infrastructure est mise en œuvre sur un réseau p2p.

**Localisation** : la recherche sous la forme de clé de hash distribuée telle qu'elle est définie dans les réseaux DHT permet de retrouver directement un composant si on en connaît son nom.

**Réplication** : la réplication sous tendue par les réseaux à traînée garantit la disponibilité des composants même en cas de panne partielle de certains nœuds

**Small-world** : dans cette approche, plus une ressource est nécessaire, plus elle est répliquée, et à l'inverse si une ressource n'est plus utilisé elle n'est plus disponible qu'à sa source.

**Identification et authentification** : l'identification et l'authentification des ressources doit être contrôlé directement par l'infrastructure. Ainsi, il ne doit pas être possible de substituer une ressource par une autre, et il faut pouvoir distinguer deux ressources dont le nom est similaire voire identique.

Notre mise en œuvre repose sur un réseau de type *pastry*, car il offre d'une part les possibilités de réplication et de DHT et d'autre part il existe une implantation open-source directement exploitable [FRE 04]. Dans cet article nous nous focalisons uniquement sur l'implantation de notre infrastructure. Les propriétés de type « petits-mondes » et celles liées à la sécurité n'y sont pas abordées. La section suivante présente notre mise en œuvre d'un réseau p2p de gestion de composants logiciels dans le cadre de la plate-forme OSGi.

## 4. Mise en œuvre d'un réseau pair-à-pair de déploiement

Notre approche répartit un ensemble de composants logiciels sur un ensemble de machines. Le rôle du réseau est de choisir la ou les machines sur lesquelles chaque composant est stocké, et de permettre de rechercher le plus rapidement possible un composant pour le fournir au client qui désire l'installer.

### 4.1 API de manipulation du réseau par l'infrastructure de déploiement

**Gestion des nœuds** : les nœuds peuvent joindre et quitter le réseau.

**Gestion des ressources** : les ressources sont indexées sur leur nom et déposées sur un ensemble de nœuds. Une ressource peut être retirée du réseau.

**Localisation et transport des ressources** : le réseau doit permettre de trouver la ressource et de la transporter jusqu'au client qui désire l'installer.

**4.2 Description du déploiement de composants OSGi**

*4.2.1 Architecture générale du réseau*

Notre infrastructure de gestion de déploiement de composants OSGi est structurée autour d'un réseau p2p *Pastry*. Il est structuré en trois couches (Figure 2) : la couche IP identifie les passerelles OSGi participantes. La couche *pastry* donne un identifiant de nœud à toutes les passerelles. La couche composants identifie la localisation des composants OSGi sur les nœuds *pastry*. Lorsque nous parlons de réseau dans la suite de l'article il s'agit du réseau de nœuds *pastry*.

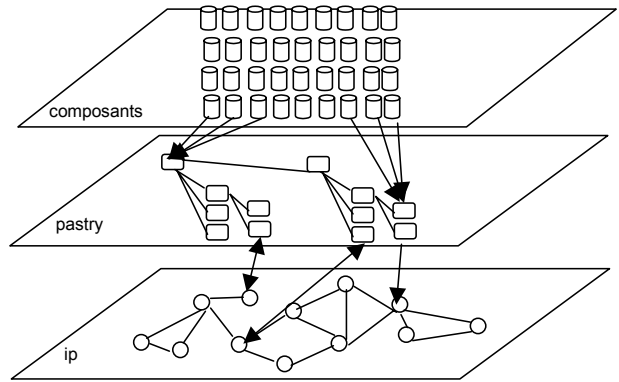

*Figure 2 : overlay de gestion du déploiement de composants*

Sur la figure 2 nous n'indiquons que la racine de dépôt d'un composant la localisation du composant est « traîné » sur le chemin entre la source du composant et la racine de stockage. La mise en œuvre de l'infrastructure p2p dans OSGi est décrite dans la suite de cette section.

*4.2.2 URI généralisée p2p://*

Un *bundle* OSGi est une archive java au format jar dont le nom sert d'identifiant. Pour installer un composant, l'utilisateur lance la commande `start <URL_du_Bundle>`. Afin de rester cohérent avec la notion d'url, nous proposons d'utiliser un mécanisme de type URI dont le schéma est p2p://<bundleName>. Ainsi la commande start p2p://log.jar installe localement un package log.jar qui est trouvé sur le réseau de pairs.

*4.2.3 Diffusion des composants*

N'importe quel pair peut mettre à disposition du réseau un composant logiciel. Pour cela il envoie une demande de dépôt de composant. Chaque nœud du réseau possède la fonction de hash qui permet de calculer le nœud destination de la ressource : Hash(log.jar) ➔ Id. Le routage entre le nœud courant qui diffuse le composant et le nœud qui va recevoir le composant log.jar est fait automatiquement par le réseau. Le réseau route le composant de passerelle en passerelle jusqu'à la passerelle racine.

*4.2.4 Insertion d'un noeud*

Un nœud s'insère dans le réseau en connaissant au moins un membre de la communauté. L'insertion se fait en utilisant les fonctionnalités du réseau *pastry*. Lors de l'insertion les nœuds se répartissent les ressources hébergées.

**4.3 Mise en œuvre dans FreePastry/Oscar**

Nous avons mis en œuvre une première implantation de notre approche. Nous nous somme fondé sur les distributions opensource Oscar [OSC 04] pour la plate-forme de services OSGi et *Freepastry* [FRE 04] pour la distribution sur le réseau pair-à-pair. Pour la mise en œuvre nous avons un composant d'interface au réseau *pastry*, un composant d'administration des composants et nous avons modifié l'outil d'installation des composants *OBR* (*Oscar Bundle Repository*).

*4.3.1 Le composant d'interface à freePastry*

Le *bundle* pastryWrapper.jar permet de gérer l'interface avec le réseau de pairs. Il est utilisé par la passerelle pour se déclarer sur le réseau et par les gestionnaires de *bundles* pour gérer les ressources.

*4.3.2 Le bundleRepository*

Le *bundle* bundleRepository de la distribution Oscar permet de gérer l'installation à distance de composants logiciels. Sur la dernière version cela est réalisé en exploitant un descripteur XML de *bundles*, qui lui permet de connaître les descriptions des *bundles* hébergés. Ce fichier est initialement récupéré par le nœud client, puis analysé afin de monter une représentation en mémoire des composants qu'il peut installer. L'installation se fait par des requêtes http à partir des url déclarées dans le fichier de description. Dans notre approche les urls sont étendues afin d'y intégrer un gestionnaire de protocole dont l'URI est de type p2p://. Ainsi si la localisation du *bundle* ou des sources du *bundle* est de type p2p://bundlename.jar, obr le recherche directement sur le réseau pair-à-pair.

*4.3.3 Le bundleRepositoryAdmin*

La déclaration des *bundles* sur le réseau p2p n'est plus réalisée sur un site centralisé. Dans ce cas, chaque machine dépose les *bundles* qu'elle veut mettre à disposition en passant par le service obrAdmin local. Ce service calcule l'identifiant du nœud racine(nœud destination du chemin de routage) puis le lui envoie.

**4.4 Synthèse**

Nous avons réalisé une implantation d'un réseau de gestion du déploiement de composant OSGi sur une infrastructure Oscar. Notre implantation repose sur trois *bundles*. (obr, obradmin, pastryWrapper). La réalisation est disponible en openSource sur http://darts.insa-lyon.fr/.

Notre approche répond aux besoins de diffusion et de localisation de *bundles* que nous avons exprimé dans la section précédente. L'infrastructure offre des fonctionnalités de stockage de composants qui sont répartis uniformément sur un ensemble de nœuds. Cette infrastructure équilibre la charge de gestion des composants au travers des participants du réseau. De plus une ressource qui n'est jamais utilisée peut disparaître du chemin de routage, tout en restant à la racine et à la source. Cette approche nous permet d'avoir une forme d'adaptation du réseau au stockage des composants.

Notre implantation du réseau repose sur une approche à composants. Ainsi l'API d'accès aux fonctions du p2p est définie dans un service OSGi et est mise en œuvre dans une implantation spécifique (dans notre cas *freepastry*). Cette approche nous permet d'être indépendant du réseau p2p sous-jacent.

Le réseau p2p est mis à disposition comme un service de déploiement. Un nœud quelconque du système qui ne lance pas le service d'administration ne participe pas au stockage des composants, mais peut toujours effectuer une recherche sur le réseau de pairs, s'il connaît un point d'entrée au réseau. Ainsi il ne contraint pas tous les pairs à participer à la gestion du déploiement.

Enfin, les ressources initialement installées sur un nœud sont conservées localement, même si un nouveau nœud en acquiert la responsabilité lors de son insertion dans le réseau. Cependant nous supposons que la communauté peut être dynamique et les nœuds qui apparaissent peuvent également disparaître. En ne supprimant pas les ressources qui sont transférées nous pouvons absorber l'aspect dynamique du réseau. Cette approche suppose que les nœuds les plus stables sur le réseau (plus présents) possèdent un grand nombre de composants.

**5 Commentaires et Conclusions**

Dans le cadre de la gestion du déploiement de composants logiciels, nous avons mis en œuvre une infrastructure de déploiement et de téléchargement de composants OSGi sur un réseau pairs-à-pairs. Notre approche soulève cependant un certain nombre de commentaires que nous exprimons dans cette section.

- Les réseaux pairs à pairs

Il existe de très nombreux choix possibles pour la mise en œuvre de notre infrastructure. Dans une première approche, nous nous sommes focalisé sur une mise en œuvre la plus simple et directe possible.

En premier lieu, nous aurions pu choisir un protocole de routage différent. Dans les travaux sur les réseaux pairs-à-pairs, il existe des spécifications de protocoles qui semblent être plus adaptés à notre problématique. Par exemple le p2p fondé sur les graphes de De-Bruijn (comme d2b [FRA 03]) semblent offrir des capacités de routage et de gestion de la communauté plus adaptés que *pastry*. Notre approche nous le permet très facilement car le *bundle* freepastrywrapper.jar déclare une interface standardisée d'accès au réseau p2p conforme aux travaux présentés dans [DAB 03].

Une solution que nous voulons étudier concerne l'utilisation simultanée de plusieurs protocoles de découverte/publication en fonction du contexte. Ainsi il serait possible d'utiliser une recherche de type broadcast sur le réseau local. Si aucune machine ne répond, on peut passer automatiquement en mode routage. Cette approche est celle proposée dans l'API *jxta* définie par Sun. Elle est intéressante dans le cas de déploiement d'applications dans des salles machines. Le téléchargement distant ne se fait qu'une seule fois. Les chargements suivants sont ensuite faits localement.

En dernier lieu, il est possible d'optimiser le transport des composants logiciels en utilisant simultanément plusieurs plates-formes de services. Cette approche similaire à la proposition de Bittorent [COH 04] ou de Pdtp [PDT 04] permet de combiner le chargement à partir de plusieurs sources de données. Cette approche n'a pas été initiée car elle suppose que les composants logiciels à installer soient d'une taille conséquente ce qui n'est pas encore notre cas.

- Les plates-formes de services

Nous avons réalisé notre implantation dans le cadre de passerelles de services OSGi. OSGi est une spécification focalisée sur la gestion du cycle de vie de composants logiciels appelés *bundle*s. Elle considère que les *bundles* sont téléchargés à partir d'une URL. Il est donc envisageable de considérer des url de type p2p://, à condition que la passerelle concernée accepte ce protocole de transport. Nous proposons d'implanter ce gestionnaire de protocole directement dans la passerelle.

Enfin les principes que nous avons présentés ne sont pas limités à la spécification OSGi ou à une distribution de *bundles*. Il est tout à fait possible de l'appliquer dans le cadre de la diffusion de logiciels comme des distribution linux par exemple. Actuellement un miroir debian nécessite 100Go pour la totalité du système. Notre approche permet de réaliser un tel miroir sur une infrastructure de plusieurs machines reliées par un réseau pair-à-pair.

## *5. Bibliographie*